\documentclass[runningheads]{llncs}
\usepackage{amsmath}
\usepackage{graphicx}
\usepackage{comment}
\usepackage{hyperref}
\usepackage{wrapfig}
\usepackage[font=scriptsize,skip=0 pt]{caption}
\usepackage{diagbox}

\usepackage{stackengine,amssymb,graphicx,scalerel}

\newcommand\CircArrowRight[1]{\stackengine{-.2ex}{\scalebox{.5}{#1}}{\CAR}{O}{c}{F}{F}{L}}
\newcommand\CAR{\scaleto{\circlearrowright}{2ex}}

\begin{document}

\title{
Interactive Process Improvement Using Simulation of Enriched Process Trees}
%\thanks{\scriptsize Funded by the Deutsche Forschungsgemeinschaft (DFG, German Research Foundation) under Germany's Excellence Strategy – EXC 2023 Internet of Production- Project ID: 390621612. We also thank the Alexander von Humboldt (AvH) Stiftung for supporting our research.}
\titlerunning {Interactive Process Improvement}
\author{Mahsa Pourbafrani \and 
Wil M. P. van der Aalst}
\authorrunning{M. Pourbafrani and Wil M. P. van der Aalst}
\institute{Chair of Process and Data Science, RWTH Aachen University, Germany \\
 \email{\{mahsa.bafrani, wvdaalst\}@pads.rwth-aachen.de}\\
 }

\maketitle              % typeset the header of the contribution
\begin{abstract}
Event data provide the main source of information for analyzing and improving processes in organizations. Process mining techniques capture the state of running processes w.r.t. various aspects, such as activity-flow and performance metrics. The next step for process owners is to take the provided insights and turn them into actions in order to improve their processes. These actions may be taken in different aspects of a process. However, simply being aware of the process aspects that need to be improved as well as potential actions is insufficient. The key step in between is to assess the outcomes of the decisions and improvements.
In this paper, we propose a framework to systematically compare event data and the simulated event data of organizations, as well as comparing the results of modified processes in different settings. The proposed framework could be provided as an analytic service to enable organizations in easily accessing event data analytics. The framework is supported with a simulation tool that enables applying changes to the processes and re-running the process in various scenarios. The simulation step includes different perspectives of a process that can be captured automatically and modified by the user. Then, we apply a state-of-the-art comparison approach for processes using their event data which visually reflects the effects of these changes in the process, i.e., evaluating the process improvement. Our framework also includes the implementation of the change measurement module as a tool.

\keywords{process mining, business process improvement, process simulation, earth mover's distance, performance spectrum.}
\end{abstract}

\section{Introduction}
\label{sec:intro}
Process owners use data-driven process mining techniques to improve their processes. The discovered process models, their performance states, and hidden problems, such as deviations and bottlenecks, are critical to process improvement. The process mining techniques in the process discovery and conformance checking areas are widely used to illustrate the current states of processes and their potential problems \cite{DBLP:books/sp/Aalst16}.
However, before taking any action based on process mining diagnostics, one wants to have an estimation of the impact. %the insights is not sufficient to change processes. The ability to quantify the actions' results is indispensable. 
To do so, it is required to play out the processes with the process owners' adjustments and then assess the effects of the actions.
%Taking actions based on the insights is not sufficient to improve the processes, the important part is to be able to measure the effects. To do so, an intermediate tool is required to play out the process behaviors with process owners' changes and afterward measure the effects of the actions.
To improve processes in an evidence-based manner, \emph{forward-looking} process mining techniques such as prediction and simulation are needed. They enable what-if and scenario-based analyses of business processes. However, the validity of the generated results, as well as their clear interpretation, are two determining factors when employing these techniques. The model's reliability can be improved by incorporating process mining insights, e.g., the designed simulation model is derived directly from the process's historical event data \cite{SummerSimKeynote2018}.

Techniques such as generating CPN models \cite{DBLP:journals/is/RozinatMSA09,CPNToolMahsa2021,PNSimMahsaArxive2020} and BPMN Models \cite{DBLP:conf/bpm/CamargoDR19} have been proposed for generating simulation models of processes based on event logs. Simulation approaches in process mining are also useful for other applications. In \cite{ConformanceCheckingSimuMohamreza2020}, for example, process model simulations are used to estimate the alignment value. The gap that we aim to fill is not only providing a platform for users to easily re-run their processes using the automatically generated simulation models but also a more accurate technique for measuring improvement/changes w.r.t. the process owners' interactions with the process.
The conventional comparison of two processes includes conformance checking between the event logs and the corresponding process models. In addition, for the purpose of performance comparison, general performance metrics are usually considered. Most of the current approaches are not detailed enough in both aspects, i.e., conformance checking and performance analysis. These techniques do not measure and reflect the effect of changes at the detailed level. For instance, the existing conformance checking techniques only return a value such as the fitness of two event logs, or one event log and the corresponding process model \cite{conformJosef}. These techniques also neglect the importance of the frequency of process instances. The detailed distance between the original event log and the regenerated event log is critical for determining their similarity \cite{MajidQuantification}. %\cite{SanderLeemanEMD}.

In this paper, we propose an approach to systematically compare the event data of a process with its simulated event data to assess the reliability of the simulation model, i.e., the accuracy of the simulation. As a result, the simulated processes in different settings can be compared. The simulation module is implemented as a new software capturing different process perspectives, in which the event logs are used to enrich the process models (trees) with existing aspects. The enriched process trees generate process behaviors in the form of event logs with/without applied changes to the process. The state-of-the-art comparison framework is then applied to the results of the simulation. It measures the effects of changes using detailed conformance and performance techniques. 
To demonstrate a proof of concept of the framework, we use a sample process as an example to illustrate the approach steps. Then,  we employ a real-life event log to evaluate the approach.

The remainder of this paper is structured as follows. 
We present the related work in Section \ref{sec:rel}.
In Section \ref{sec:background}, we introduce background concepts and notations.
In Section \ref{sec:approach}, we present our main approach. We evaluate the approach in Section \ref{sec:eval} by designing simulation models, and Section \ref{sec:conclusion} concludes this work.

\section{Related Work}
\label{sec:rel}
Process mining enables designing data-driven simulation models of processes \cite{SummerSimKeynote2018}. Authors in \cite{DBLP:journals/is/RozinatMSA09} use different aspects of a process using its event data, e.g., process models, resource pooling, and performance metrics, and automatically generate simulation models. This work as a pioneer in the data-driven simulation in process mining translates insights from event data into the process simulation parameters. Other simulation approaches in process mining follow the same direction. For instance, \cite{DBLP:journals/is/Rogge-SoltiW15} uses \emph{stochastic Petri nets} to simulate processes and determine the duration of instances in business processes. In \cite{DBLP:conf/bpm/PufahlW17} a business simulation model is generated which is based on the user domain knowledge. Tools based on \emph{Protos} try to reduce modeling efforts by introducing the reference process models \cite{protosVerbeek}. \cite{martin2016use} discusses how process mining insights can be exploited in the business process simulation context. As an example, the proposed tool in \cite{DBLP:conf/bpm/CamargoDR19} presents the idea of combining BPMN and process mining for simulation purposes, where indicators for measuring the accuracy of the simulation results are also introduced. 
%\cite{DBLP:journals/corr/abs-1910-05404} 

In \cite{FeatureExtractionCaise21,mahsaTimeseries}, different levels of simulating processes are proposed where all the aspects of a process are extracted at different levels, i.e., not only instance level but also higher-level, e.g., describing processes per day quantitatively.
The examples of high-level simulations are presented in \cite{MahsaBIS,DBLP:conf/ihsi/PourbafraniZA20} with the use of the designed tool for the modeling and data extraction steps in \cite{mahsaToolPMSD}.
In our approach, the enriched process models, e.g., process trees, accuracy of the performance-related aspects, effortless interaction with users, and social network analysis (resource aspects) are the main criteria for designing simulation models. 

On the other side, visualization techniques are powerful tools in process mining analysis in both descriptive and predictive analyses. There are a couple of visualization techniques that are able to represent the process w.r.t. different process aspects for providing visual inspection or process comparison. For instance, the performance spectrum \cite{PerformanceSpecVadim2018} represents the process performance behaviors in detail between every two sets of activities in the process. i.e., process segments. %Moreover, for the conformance checking techniques, the deviation representation on top of the process models is a common practice \cite{PerformanceConformance2012Wil}. 
The stochastic conformance checking method used in \cite{MajidToolEMD} considers the frequency of the traces in two event logs while comparing their differences. The idea of using \emph{Earth Mover's Distance} for conformance checking and comparing two event logs, or event logs and process models enables assessing the difference of two processes w.r.t. their behaviors in detail. 

%In our approach, we use enriched process trees for automatically regenerating the process behaviors and innovative visualization methods to measure the accuracy of the simulated models, and the improvement of the applied changes. 

We provide a platform for regenerating a process in different settings and measure the effects of changes/results using our designed modules based on the presented ideas. The presented tool in \cite{MahsaToolDemoPTree2021} is the simulation approach taken in the current work as the intermediate tool for regenerating the process behaviors. The process trees are automatically generated and enriched with the probability and performance information and allow us to change the processes w.r.t. the activity-flow and performance aspects.

\section{Preliminaries}
\label{sec:background}
In this section, we establish the basic notations for events, event logs, and process trees which are used in the framework. 
\begin{definition}[Event]
\label{def:eventlog}
Let $\mathcal{A}$ be the universe of activities, $\mathcal{T}$ be the universe of timestamps, $\mathcal{R}$ be the universe of resources, and $\mathcal{C}$ be the universe of case identifier. An event e is a tuple $e{=}(c,a,r,t)$ where activity $a$ at time $t$ for case $c$ is performed by resource $r$. $ \mathcal{E} {=} \mathcal{C}\times\mathcal{A}\times\mathcal{R}\times\mathcal{T}$ is the universe of events. For each $e\in \mathcal{E}$, $\pi_{D}(e)$ projects $e$ on the attribute from domain $D$, e.g., $\pi_{\mathcal{A}}(e){=}a$.

\end{definition}
%\begin{wraptable}{r}{0.48\textwidth}
\begin{table}[bt]
\centering
%\vspace{-25 pt}
\caption{A part of a sample event log. Each row represents an event.} %For instance, Peter registered a request for the customer with id 1 at timestamp .  }
\tiny
\label{table:exmeventlog}
\resizebox{0.7\textwidth}{!}{

\begin{tabular}{l|c|c|c|c|}
\cline{2-5}
      & Case ID & Activity           & Resource & Timestamp        \\ \cline{2-5} 
$e_1$ & 1       & register request   & Pete     & 12/30/2010 11:02 \\ \cline{2-5} 
$e_2$ & 2       & register request   & Mike     & 12/30/2010 11:32 \\ \cline{2-5} 
      & 3       & register request   & Pete     & 12/30/2010 14:32 \\ \cline{2-5} 
...   & 1       & examine thoroughly & Sue      & 12/31/2010 10:06 \\ \cline{2-5} 
      & 2       & decide             & Sara     & 1/5/2011 11:22   \\ \cline{2-5} 
      & 1       & decide             & Sara     & 1/6/2011 11:18   \\ \cline{2-5} 
      & 1       & reject request     & Pete     & 1/7/2011 14:24   \\ \cline{2-5} 
$e_n$ & ...     & ...                & ...      & ...              \\ \cline{2-5} 
\end{tabular}
}
\end{table}
%For instance, each row in \autoref{table:exmeventlog} is an event in the running example. Consider the first row as $e_1$ in which for the customer with case id $1$ ($\pi_{\mathcal{C}}$), activity \emph{register request} ($\pi_{\mathcal{A}}$) is performed by $Pete$ ($\pi_{\mathcal{R}}$) at $12/30/2010 11:02$ ($\pi_{\mathcal{T}}$).

\begin{definition}[Trace]
Let $\mathcal{E}$ be the universe of events, a trace $\sigma{\in}{\mathcal{E}^*}$ is a finite sequence of events.
For each $\sigma{=}{\langle}e_1, ..., e_n{\rangle}$, $e_{i}{\in}\sigma$ happens at most once and for each $e_i,e_j \in \sigma, \pi_C(e_{i}){=}\pi_C(e_{j})\wedge\pi_{\mathcal{T}}(e_i){\leq}\pi_{\mathcal{T}}(e_j), \textit{if } 1\le i < j \le n$. For $\sigma{=}{\langle}e_1, ..., e_n{\rangle}{\in}\mathcal{E}^*$, $\Pi_D(\sigma)=\langle \pi_{\mathcal{D}(e_1)},\pi_{\mathcal{D}(e_2)}, ... , \pi_{\mathcal{D}(e_n)} \rangle$ is the projection of trace $\sigma$ on the attribute from domain $D$, e.g., $\Pi_{\mathcal{A}}(\sigma)=\langle \pi_{\mathcal{A}(e_1)},\pi_{\mathcal{A}(e_2)}, ... , \pi_{\mathcal{A}(e_n)}\rangle$.  
\end{definition}
 %and $a,f,c,e$ is another behavior which only happened for $10\%$ of customers. 

\begin{definition}[Event Log]
Let $\mathcal{E}$ be the universe of events and $\mathcal{E}^*$ be the set of possible traces, we define an event log $L$ as a set of traces, i.e.,  $L\subseteq \mathcal{E}^*$. 
\end{definition}

We denote $L_{\mathcal{A}}{=}[\Pi_{\mathcal{A}}(\sigma)|\sigma \in L]$ as the multiset of traces projected on the activity attribute. Furthermore, $\widetilde{L_{\mathcal{A}}}{=}\{\sigma\in L_{\mathcal{A}}\}$ is the set of unique traces (variants) projected on the activity attribute in the event log $L$. We refer to $\widetilde{L_{\mathcal{A}}}$ as the set of process behaviors presented in $L$.

%We denote $\widetilde{L}{=}\{\sigma\in L\}$ as the set of unique traces in the event log $L$. We refer to each trace as behavior that the process represents. For instance, in the running example, the sequences of activities, \emph{register request, examine thoroughly, check ticket, decide and reject request}, performed for $40\%$ of costumers is one of the process behavior (trace).
\begin{wrapfigure}{r}{0.48\textwidth}
\vspace{-7 mm}
    \centering
    \includegraphics[height=0.1\textheight,width=0.5\textwidth]{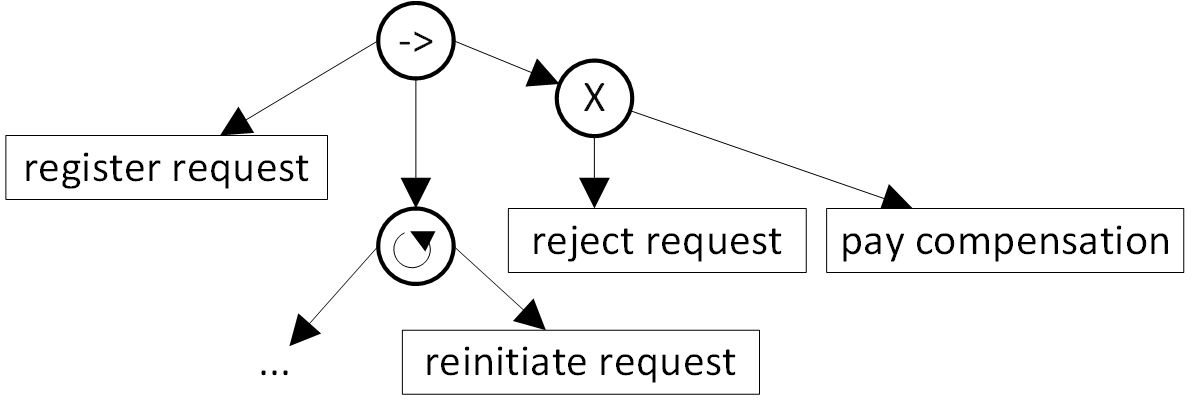}
    \caption{A part of the discovered process tree for the sample event log. }
    \label{fig:Ptreeexample}
    \vspace{-7 mm}
\end{wrapfigure}
\autoref{table:exmeventlog} represents a part of a sample event log, where each row indicates an event, e.g., considering the first row as $e_1$, $\pi_{\mathcal{C}}(e_1)=1$ and $\pi_{\mathcal{A}}(e_1)= register\ request$. Process mining utilizes such event logs to discover running processes inside organizations. 
The process models are the representative ways of the discovered running processes. The process tree notation is one of the common approaches to present a process, where the nodes of trees are operators and leaves are activities in the process. 

A part of the process tree representing the example process is shown in \autoref{fig:Ptreeexample}. For example, there is a choice, i.e., $XOR$ ($\times$) as a node between activity $reject\ request$ and $pay\ compensation$ indicating that in the process either a request is rejected or the compensation is paid. The root node ($\to$) indicates that activity $register\ request$ is always followed by a loop (\CircArrowRight{}). A loop represents a redo of works between its children, i.e., activities in the leaves of a loop node may happen multiple times in a trace. Furthermore, the notation of $\tau$ is for silent activities which are not visible in the process but used for the representation of process trees. 
\begin{definition}[Process Tree]
\label{def:processTree}
Let $L$ be an event log, $A_L=\{a\in \sigma |\sigma \in \widetilde{L_{\mathcal{A}}}\}$ be the set of activities in $L$ and $Op{=}\{\to, \times, \CircArrowRight{}, +\}$ be the set of process operators.
If $a\in A_L \cup \{\tau\}$, then $Q=a$ is a process tree. If $n\geq 1$, $Q_1,Q_2,Q_3,...,Q_n$ are process trees, and $op\in \{\to, \times, +\}$, then $Q=op(Q_1,Q_2,...,Q_n)$ is a process tree. If $n\geq 2$ and  $Q_1,Q_2,Q_3,...,Q_n$ are process trees, then $Q=\CircArrowRight{}(Q_1,Q_2,...,Q_n)$ is a process tree. 
For a process tree $Q$, we denote $Q_a$ and $Q_{op}$ as the set of activities and the set of operators in $Q$. 
\end{definition}
For a given process tree $Q$, $Q_w=Q_{op}\times Q_{a}$ is the set of edges connecting operators to activities. For instance, $(\rightarrow, register\ request)$ is an edge in the example process tree in \autoref{fig:Ptreeexample} where $register\ request$ is child of the tree under parent $\rightarrow$. Note that a process tree may also contain edges from an operator to an operator, which is not relevant in the implementation of our framework.

  %is followed by a  are parallel and ... is sequential, x is always by y in the process.

\section{Approach}
\label{sec:approach}

%Measure the changes in the process behavior (activity-flow, performances) in the following scenarios: The same process with the extracted insights from the event log The changes in conformance and performance aspects of the process, e.g., added new behaviors, by Changing operators in the process tree Changing the performance aspects of the process tree

\begin{figure}[bt]
    \centering
    \includegraphics[width=0.9\textwidth,height=0.28\textheight]{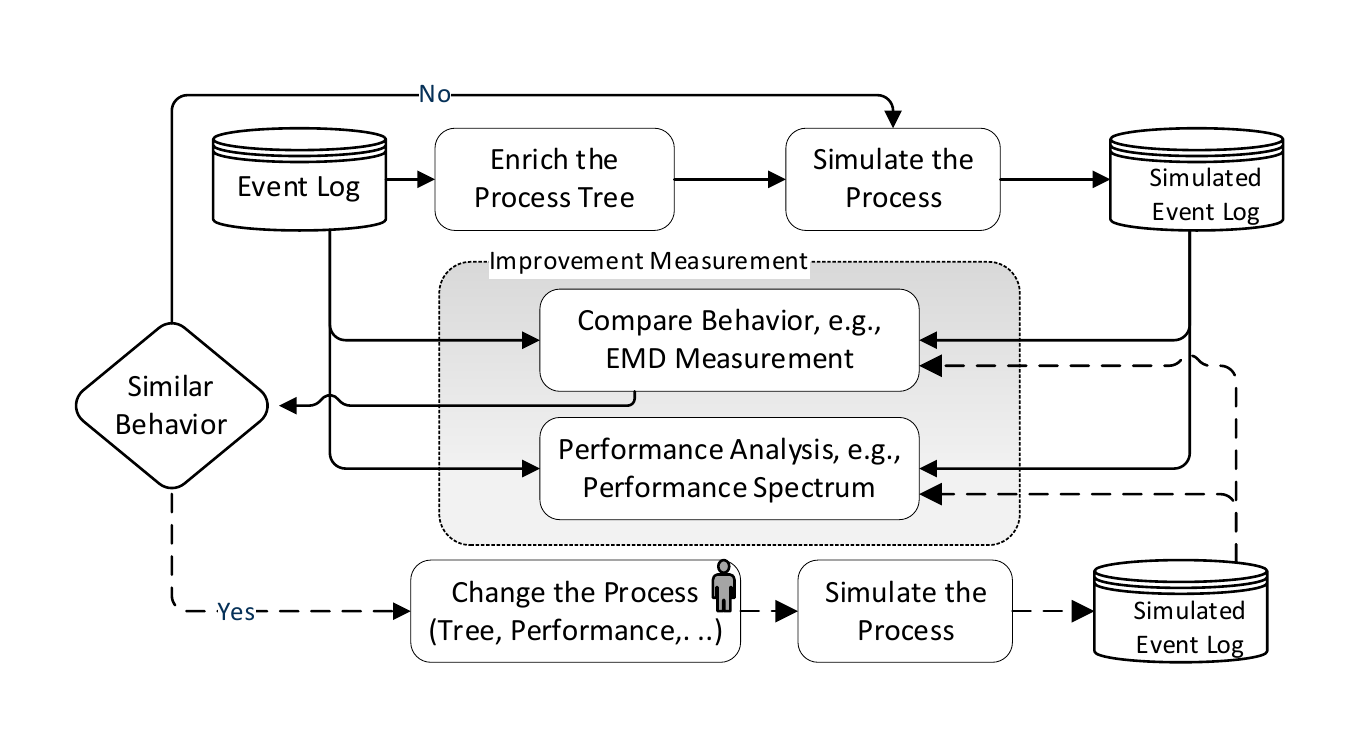}
    \caption{The overview of the framework to improve the processes interactively. %Using event logs and process mining techniques the setting for the simulation models as an intermediate tool is automatically performed. 
    The straight lines show the path to assess the quality of the regenerated behavior by the simulation model w.r.t. activity-flow and performance metrics. The dotted lines illustrate the path that the user is able to change the process and measure and observer the improvement, i.e., the effect of changes, in the process. }
    \label{fig:ImprovePTFramework}
\end{figure}

Our framework enables interactive process improvement inside organizations for designing/improving process models. The current behaviors of processes captured in the form of an event log serve as the starting point for any improvement. To enrich the discovered process models, process discovery, performance analysis, and social network analysis (resource perspective) techniques are used. We use the \emph{Discrete Event Simulation} (DES) technique as a tool to play out the process with the current states, which results in an event log as shown in \autoref{fig:ImprovePTFramework}. The original behavior of the event log w.r.t. activity-flow (process behavior) and performance metrics are compared to ensure that the automatically designed simulation model is reliable and behaves close to reality, \emph{Improvement Measurement} module in \autoref{fig:ImprovePTFramework}. This step allows the user to change the process parameters and re-run the process to generate the new behavior and measure the process improvement, depicted by the dotted lines in \autoref{fig:ImprovePTFramework}. These measurements are presented in a numerical format as well as in a detailed graphical format. The detailed comparative visualization increases the interaction between the framework and the user. First, we explain the automatic generation of the simulation results, including process mining techniques and enriching the process model, and continue with the \emph{Improvement Measurement} module.  %As the next step, the simulation parameters of the generated models can be configured and executed. Hereafter, the two major measurements are behavior comparison and performance comparison which illustrate the effect of changes in the process.

\subsection{Simulating Process Trees}

\subsubsection{Enriching Process Trees}
%Measuring the difference between the actual process behavior represented in the event log, Therefore, 
The inductive miner algorithm \cite{inductiveminerSander} is used to discover the process model since it is capable of capturing all the behaviors in a process in the form of a process model. %The inductive miner algorithm with the noise parameter of $0$ generates process trees with a fitness value of 1 indicating that the process tree is able to represent all the traces in the event log. 
The generated process tree by the inductive miner algorithm is able to represent the traces in the event log. 
The process tree's limited number of operators as defined in \autoref{def:processTree} allows for easy understanding and modification of the process. To play out the process accurately, i.e., applying the new changes in the process, more information than the flow of activities provided by the process tree is required. 

The tree should be enriched with the probability of activity-flows, performance information of the activities, and the corresponding resource information, e.g., organizations of the resources, the number of resources in each organization, and hand-over of activities between resources, for each activity from the real process. Therefore, the probability of the choices and the possible number of loops should be taken into account for regenerating a similar event log. 
Furthermore, for a process tree $Q$ and the edge $w=(op,a)\in Q_w$, $w_a$ represents the probability of occurrence of activity $a$ in a generated trace from the process tree. For the edge $w=(op,a)\in Q_w$, if $op\in\{\rightarrow, +, \CircArrowRight{}\}$, then $w_a=1$.  %and if $op=\times$, then $w_a$ is the probability of occurrence of $a$ in a trace. 
Note that to avoid the generation of infinite traces due to the loops in the process tree, we limit the execution of loops in the simulation with the probability of the number of occurrences of a loop on average in a trace and the maximum times that a loop happens in a trace.
%Furthermore, if $Op=\{\times\}$, then for all $a \in Q$, there is a bounded weight $w_a$ which represents the probability of happening. If $Op=\{\CircArrowRight{}\}$, then for all $a \in Q$, there is a bounded weight of $w_a$ representing the probability of the number of occurrence of a loop on average in a trace and the maximum times that a loop happens in a trace. For $Op=\{+\}$ or $Op=\{\longrightarrow\}$ the probability of occurrence of activities involved in the tree is always $1$. 
For all activities $a\in Q_a$, there is a binding performance metric, i.e., the average duration of each activity. Moreover, the activities are assigned to the existing automatically discovered organizations and the capacity of the resources.

%To do so, we define probability aware-process trees in \autoref{def:probProcessTree}.

%\begin{definition}[Probability-aware Process Tree] \label{def:probProcessTree}Let $Q$ be a process tree. If $Op=\{\times\}$, then for all $a \in Q$, there is a bounded weight $w_a$ which represents the probability of happening. If $Op=\{\CircArrowRight{}\}$, then for all $a \in Q$, there is a bounded weight of $m_a$ representing the maximum happening loop in a trace. \end{definition}
\begin{figure}[bt]
    \centering
    \includegraphics[height=0.25\textheight]{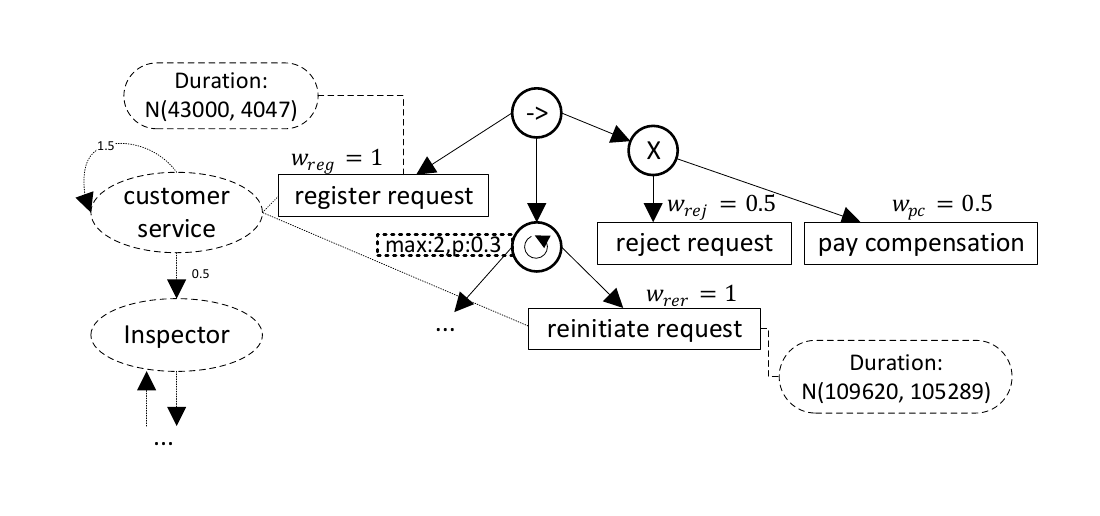}
    \caption{Enriched tree with the probability information, resource allocation, and duration of each activity. The enriched process tree can be simulated. The hand-over of resources is shown (left) to provide more accurate simulation results (event logs) w.r.t. the resource allocation in the organizations.} %represented using their organization and the probability of had over the work from one organization to others.  }
    \label{fig:EnrichedProcessTreeRunExmp}
\end{figure}
For the example process shown in \autoref{fig:Ptreeexample}, a part of the automatically enriched tree with the activity-flow, performance, and resource information is presented in \autoref{fig:EnrichedProcessTreeRunExmp}. For instance, for the process edge $w=(\times, reject\ request)$, $w_{rej}= 0.5$, and the shown loop in the process can be executed at most $2$ times in a trace and the probability of its occurrence is $30\%$ which is derived from the event log. Activity $register\ request$ takes on average $43000$ seconds to be performed, and the average is used for simulating its duration using a normal distribution. Also, $register\ request$ and $reinitiate\ request$ belong to the $customer\ service$ organization where the resources in this organization hand over tasks to the $inspector$ organization. 
%$\{max:2,p:0.3\}( \to( +( check\ ticket(1),\times( examine\ casually(0.67), examine\ thoroughly(0.33))),\\decide), reinitiate\ request)(1), \times(pay\ compensation(0.5), reject\ request(0.5))$. The information for the loop node specifies that the probability of the occurrence of the loop in a trace is $30\%$ and the maximum times that a loop happens in a trace is $2$ times.

The information extracted from event logs is shown in \autoref{table:param}. This information, along with the discussed information for enriching process trees are the required simulation parameters. Moreover, the changeable aspects for process improvement by the user in the simulation step are specified in detail.  %the process mining insights also inserted into the model which can be changed by the user. 
The discovery and design of the simulation models including generating event logs as a result of the simulation models are represented in detail in \cite{MahsaToolDemoPTree2021}.
%We use the discovered process model and provide an interface for the user to design a new process model including all the performance and environmental attributes, e.g., changing business hours, or resource capacity. All of these changes are supported by the information derived from event logs of processes using process mining techniques as shown in \autoref{fig:Main_Framework_PMSIM}. 
\begin{comment}
\begin{figure}[bt]
    \centering
    \includegraphics[width=0.8\textwidth,height=0.17\textheight ]{InkedSimulationTreePaper_LI.jpg}
    \caption{Caption}
    \label{fig:my_label}
\end{figure}
\end{comment}
%\subsubsection{Re-run Process Trees}
\begin{table}[bt]
\caption{The general list of automatically discovered insights using process mining techniques to form process simulations. The top row shows what is discovered from event data. The bottom row shows what can be set or change by the user.} %The changeable parameters by users in the process where the user can regenerate the process with new parameters and measure improvement/effect of changes are marked with $+$.} %the All the parameters are discovered using process mining from event logs by default are filled with the real values automatically and the execution values guaranteed the default values in case that users do not change the parameters.}%in the tool derived from process mining along with the required execution parameters. }
\label{table:param}
\resizebox{\textwidth}{!}{
\begin{tabular}{c|c|c|c|c|c|c|c|c|c|c|c|c|}
\cline{2-13}
 &
  \multicolumn{10}{c|}{Process Mining} &
  \multicolumn{2}{c|}{\begin{tabular}[c]{@{}c@{}}Simulation Execution\\ Parameters\end{tabular}} \\ \cline{2-13} 
 &
  \begin{tabular}[c]{@{}c@{}}Process\\  Model \\ (Tree)\end{tabular} &
  \begin{tabular}[c]{@{}c@{}}Arrival \\ Rate\end{tabular} &
  \begin{tabular}[c]{@{}c@{}}Activity\\  Duration,\\ Deviation\end{tabular} &
  \begin{tabular}[c]{@{}c@{}}Activities\\  Capacity\end{tabular} &
  \begin{tabular}[c]{@{}c@{}}Activities Unique \\ Resources\\ (Shared\\  Resources)\end{tabular} &
  \begin{tabular}[c]{@{}c@{}}Waiting \\ Time\end{tabular} &
  \begin{tabular}[c]{@{}c@{}}Business \\ Hours\end{tabular} &
  \begin{tabular}[c]{@{}c@{}}Activity-flow \\ Probability\end{tabular} &
  \begin{tabular}[c]{@{}c@{}}Process\\  Capacity \\ (cases)\end{tabular} &
  \begin{tabular}[c]{@{}c@{}}Interruption \\ (Process, Cases,\\  Activities)\end{tabular} &
  \begin{tabular}[c]{@{}c@{}}Start Time \\ of Simulation\end{tabular} &
  \begin{tabular}[c]{@{}c@{}}Number\\  of Cases\end{tabular} \\ \hline
\multicolumn{1}{|c|}{\begin{tabular}[c]{@{}c@{}}Automatically \\ Discovered\end{tabular}} &
  + &
  + &
  + &
  + &
  + &
  + &
  + &
  + &
  + &
  + &
  - &
  - \\ \hline
\multicolumn{1}{|c|}{\begin{tabular}[c]{@{}c@{}}Changeable \\ by User\end{tabular}} &
  + &
  + &
  + &
  + &
  + &
  + &
  + &
  + &
  + &
  + &
  + &
  + \\ \hline
\end{tabular}
}
\end{table}

\subsection{Measuring the Process Improvement}
To measure the changes in the newly generated process represented with an event log, we have to compare two event logs. For comparing two processes, i.e., event logs, two major aspects of the processes should be considered, activity-flow which generates the behaviors, and the performance aspects. %These comparisons as indicated in \autoref{fig:ImprovePTFramework} are able to measure the similarity of the generated process using simulation with the real process as well as measuring the improvement of the processes after applying the changes.
Note that the intermediate regenerator tool can be different from the one that we use in our framework, and yet the \emph{Measuring the Process Improvement} module can be used for measuring the effect of changes in two event logs.

\subsubsection{Activity-flow Behaviors}
The fact that process trees include silent transitions, loops, and $XOR$ operators makes generating more behavior (new traces) than the existed ones in the original log possible. Therefore, the similarity of behaviors is one of the main indicators in the comparison step.  
%Conformance aspect: Earth Mover Distance of two logs (original and the newly generated) If the general performance differs: Behavior of the added behaviors Behavior of the removed behaviors

Given two event logs, the original event log $L$ and the simulated event log $L'$, we show the presented behaviors in each event log using their set of unique traces, i.e., $\widetilde{{L}}_{\mathcal{A}}$, $\widetilde{{L'}}_{\mathcal{A}}$. 
%The union of two sets provides the set of all possible behaviors that happened at least in one of the logs $\widetilde{{L}}_{\mathcal{A}}\cup\widetilde{{L'}}_{\mathcal{A}} $.
%\{ \sigma | \sigma \in \widetilde{L} \vee \sigma \in \widetilde{L'} \}$. Having the union set of the behaviors, t
The new generated behaviors in the simulated event log, i.e., not existing in the original event log, and the removed behaviors from the original process are calculated as $\widetilde{{L'}}_{\mathcal{A}}\setminus \widetilde{{L}}_{\mathcal{A}}$, and $\widetilde{{L}}_{\mathcal{A}}\setminus \widetilde{{L'}}_{\mathcal{A}}$, respectively. 
%We consider the rel of the new behaviors to be 
Therefore, $\frac{|\widetilde{{L'}}_{\mathcal{A}}\setminus \widetilde{{L}}_{\mathcal{A}}|}{| \widetilde{{L}}_{\mathcal{A}}\cup \widetilde{{L'}}_{\mathcal{A}}|}$ and $\frac{|\widetilde{{L}}_{\mathcal{A}}\setminus \widetilde{{L'}}_{\mathcal{A}}|}{| \widetilde{{L}}_{\mathcal{A}}\cup \widetilde{{L'}}_{\mathcal{A}}|}$ are the fraction of the new and removed behaviors, respectively. 
%to be $\frac{\widetilde{{L}}_{\mathcal{A}}\setminus \widetilde{{L'}}_{\mathcal{A}}}{Union (L,L')}$. 
%$NewBehavior{=} \{\sigma | \sigma \in \widetilde{L'} \wedge \sigma\notin \widetilde{L}\} $, and $RemovedBehavior{=} \{\sigma | \sigma \in \widetilde{L} \wedge \sigma\notin \widetilde{L'} \} $, respectively. 
 %$Diff{=} \{\sigma| \sigma \in \widetilde{L} \vee \sigma \in \widetilde{L'} \wedge \sigma\notin Union (L,L') \}$

These metrics represent the pairwise difference between two event logs. They evaluate whether the simulation of the original log is close to reality, as well as capturing any different behavior added/removed due to the changes in a process tree (flow of activities). In the example process presented in \autoref{fig:Ptreeexample}, after regenerating the process without any change multiple times, on average $22\%$ of the generated variants (unique traces) in the simulated logs are newly generated. The sample event log for further experiments with the tools is publicly available\footnote{\scriptsize https://github.com/mbafrani/VisualComparison2EventLogs}.
%the percentage of the removed behavior in the simulation of the exact process (without any changes) is zero and $22\%$ of the generated variants (unique traces) in the simulated logs are newly generated. 
Furthermore, the precise comparison of two event logs should be based on their behavior, taking into account the frequency of the behavior. To determine the difference between the original and the simulated event logs, we employ a stochastic conformance checking approach.

\paragraph{\textbf{Earth Mover's Distance Conformance Checking}}

\begin{table}[bt]
\centering

\caption{A sample example of EMD measurement for two event logs \cite{MajidQuantification}. The reallocation function allocates the $49$ traces in $L$ to $49$ traces with activity-flow $\langle a,e,c,d \rangle$ and $1$ remaining trace to $\langle a,b,c,d \rangle$ in $L'$. The sum of the value of the table indicates the general EMD value, i.e., the difference between the two event logs. Each cell represents the minimum cost to map its corresponding trace in the original event log (row) into the traces in the simulated event log (column).}
\label{table:EMDExmaple}
\begin{tabular}{|l|l|l|l|l|} 
\hline
\diagbox{$L_{\mathcal{A}}$}{$L_{\mathcal{A}}'$}      & $\langle a,b,c,d\rangle$  & $\langle a,c,b,d\rangle$  & $\langle a,e,c,d\rangle^{49}$  & $\langle a,e,b,d\rangle^{49}$   \\ 
\hline
$\langle a,b,c,d\rangle^{50}$  & $\frac{1}{100} \times 0$                        & $0 \times 0.5$                         & $\frac{49}{100} \times 0.25 $                     & $0 \times 0.5$                          \\ 
\hline
$\langle a,c,b,d\rangle ^{50}$  & $0 \times 0.5$                         & $\frac{1}{100} \times 0$                         & $0 \times 0.5$                         & $\frac{49}{100} \times 0.25$                      \\
\hline
\end{tabular}
\end{table}
To accurately compare two event logs' behaviors, we use the probability distance of each two traces in two event logs based on Earth Mover's Distance (EMD). 
%For two piles of earth, EMD represents the amount of effort to transform one pile into the other. 
To calculate the EMD measurement between two event logs, we use the conformance techniques presented in \cite{SanderLeemanEMD}. 
For every trace in the original log, we calculate the movement of its frequency to all the traces in the simulated event log using the reallocation function. As the next step, the cost of the movement is considered using the trace distance function. 

\paragraph{Reallocation}
Let $L$ and $L'$ be the original and the simulated event logs, respectively. Function $r \in \widetilde{{L}}_{\mathcal{A}} \times \widetilde{{L'}}_{\mathcal{A}} \to [0,1]$ returns the relative frequency of $\sigma \in \widetilde{L}_{\mathcal{A}}$ that should be transformed to ${\sigma}' \in \widetilde{L'}_{\mathcal{A}}$, i.e., $r(\sigma,\sigma')$.
%the movement frequency between two event logs is calculated. The relative frequency of $\sigma \in \widetilde{L}$ that should be transformed to ${\sigma}' \in \widetilde{L'}$ is $r(\sigma,{\sigma}')$.
Note that for all $\sigma \in \widetilde{L}_{\mathcal{A}}$, $\frac{L_\mathcal{A}(\sigma)}{|L_\mathcal{A}|}{=}\sum_{\sigma' \in \widetilde{L'}_{\mathcal{A}}}r(\sigma,\sigma')$, i.e., the frequency of each $\sigma \in \widetilde{L}_\mathcal{A}$ is considered properly. The same should be considered for each $\sigma' \in \widetilde{L'}_{\mathcal{A}}$. %The reallocation function in \autoref{table:EMDExmaple} for the first trace in $L$ returns $\frac{}{}$

\paragraph{Trace Distance}
The distance between each two traces in the original log and the simulated logs is calculated based on \emph{normalized string edit distance} (Levenstein) \cite{levenshtein1966binary}. Function $d \in \mathcal{A}^* \times \mathcal{A}^* \to [0,1]$ calculates the distance between two traces, where for two similar traces the value is $0$ and $d(\sigma,\sigma'){=}d(\sigma',\sigma)$.

To represent the algorithm clearly, we reduced the sample process and presented a couple of traces in \autoref{table:EMDExmaple}. The EMD measurement of the two event logs is \sloppy{$EMD(L_{\mathcal{A}},L'_{\mathcal{A}})= \underset{r\in R}{\min} \text{ }r.d=\sum_{\sigma \in \widetilde{L}_{\mathcal{A}}} \sum_{\sigma' \in \widetilde{L'}_{\mathcal{A}}} r(\sigma,\sigma')d(\sigma,\sigma')$} where $R$ is the universe of reallocation functions. \autoref{table:EMDExmaple} represents a sample EMD measurement for two sample event logs $L$ and $L'$. For instance, for $\langle a,b,c,d \rangle^{50}$ in $L$ and $\langle a,e,c,d \rangle^{49}$ in $L'$, the trace distance value is $0.25$ given the differences between two traces using \emph{normalized string edit distance} (Levenstein). The reallocation value is $0.49$, i.e., $49$ of $100$ traces in $L$ are reallocated to $49$ traces with the sequence $\langle a,e,c,d \rangle$ in $L'$. Therefore, the minimum effort of mapping the one trace to the second one is $0.49*0.25=0.122$.
Besides the EMD value of two event logs that indicates how two event logs are different, we are interested in the required effort for every trace in the original event log to be mapped/transformed into the simulated event log for accurate comparison of the simulation results.

\begin{figure}[bt]
    \centering
    \includegraphics[width=\textwidth,]{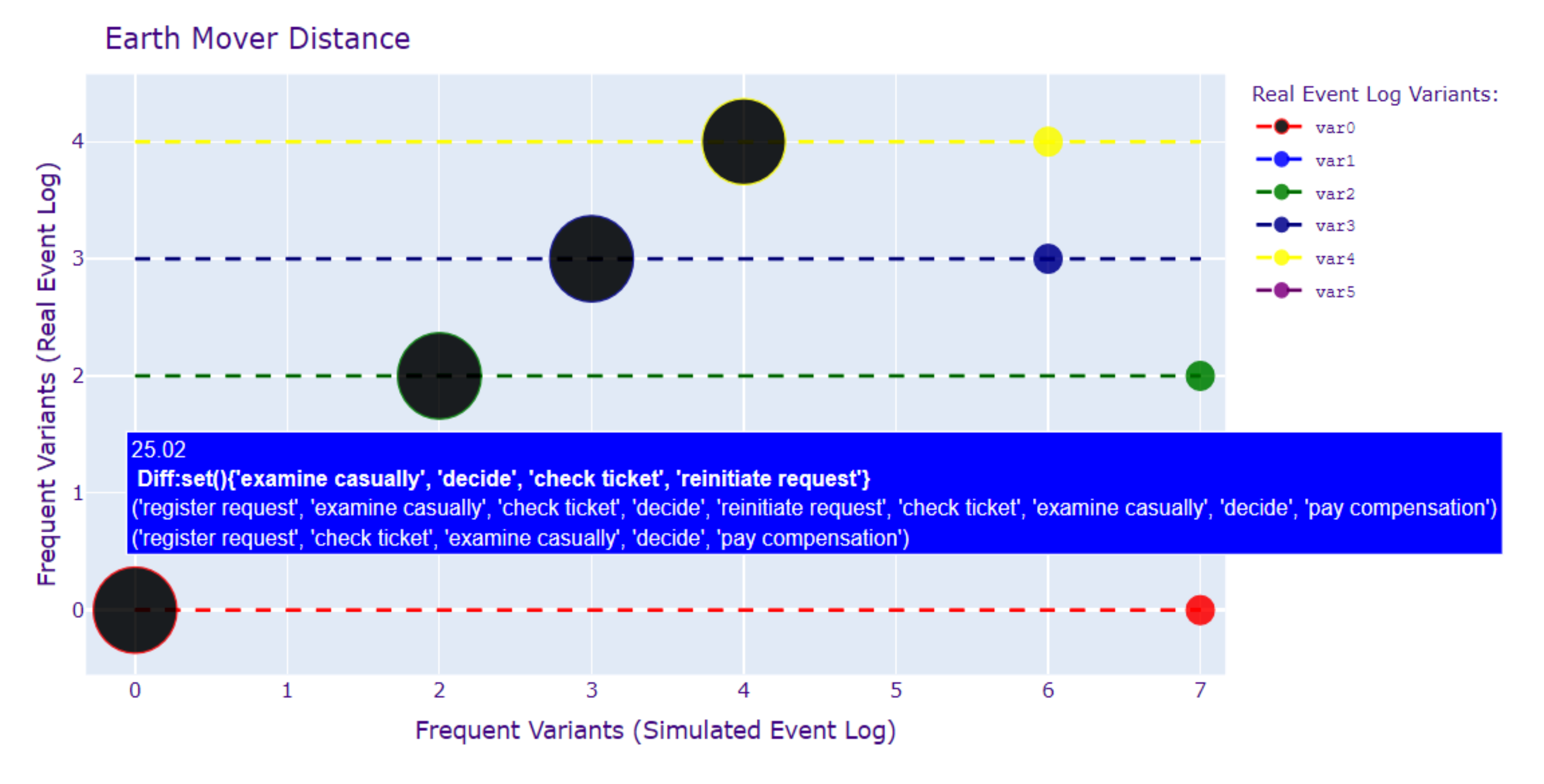}
    \caption{The detailed comparison of two event logs for the sample process, i.e., the results of EMD measurement. It is the results of the EMD \emph {reallocation} and \emph{trace distance} functions in the form of a table such as \autoref{table:EMDExmaple}. The points are the proportional cost of moving every trace in one event log to the simulated event log. Each row (color) indicates a trace in the original event log. The black points are similar traces in both event logs. The sizes of the points are the relative costs of movement for each variant (unique traces) in the original event logs. }
    \label{fig:EMDExample}
\end{figure}

Applying the designed EMD measurement to the complete sample process and its simulated event log without any changes, \autoref{fig:EMDExample} illustrates the result. The unique traces in the original event log and the unique traces in the simulated event log are depicted using the x-axis and the y-axis, respectively. If we assume that in our example, $r\in R$ is the reallocation function, the cost of EMD (effort of mapping) for each point of \autoref{fig:EMDExample} shows the relative effort, i.e., $\sigma\in \widetilde{L}_{\mathcal{A}}$ and for each  ${\sigma'_i} \in \widetilde{L'}_{\mathcal{A}}$, $\textit{effort}_{L_{\mathcal{A}},L_{\mathcal{A}}}(\sigma,\sigma'_i)=\frac{d(\sigma,\sigma'_i).r(\sigma,\sigma'_i)}{\sum_{\sigma' \in \widetilde{L'}d(\sigma,\sigma').r(\sigma,\sigma')}}$.
%Each row is the relative effort that the first unique trace in the original log need to be transformed to the the one or more unique traces in the simulated log. The value of each point in each row is calculated as follows. 
The most frequent trace in the original event log (first row) will be converted to the $( 74.98\%, 0, 0, 0, 0,0,0,25.02\%)$, i.e., points in the first row. The values indicate that to map the first trace in the original event log (most frequent one) to the simulated event log $74.98\%$ of the effort is to map it to the first (most frequent trace) in the simulated event log, i.e.,$\textit{effort}_{L_{\mathcal{A}},L_{\mathcal{A}}}(\sigma_1,\sigma'_1)=75\%$. %\autoref{fig:EMDExample} represents the detailed comparison of two event logs (running example and the regenerated one) which provides a comprehensive view of the conformance checking between them.
Also, each row illustrates the minimum required effort to map/transform the traces into the simulated event log.

\subsubsection{Performance Behaviors}
Performance is the second factor to consider when assessing improvement/changes. However, because the times are abstracted from the real data in prediction and simulation techniques, exact measurements are impossible. It is worth noting that in many cases, time-related parameters such as the duration of simulation events are generated using a random function, e.g., normal distribution in our case.  %For example, the changes in the efficiency of the resources in performing tasks. 
General performance KPIs at a high level of aggregation, e.g., the average waiting time of traces, or average service time are too abstract to represent the effects of the changes in the process. Therefore, besides the usual metrics, we use the performance spectrum, which relies on the structure of the process and directly reflects the effects of changes in specific parts of the process on others. For instance, changing the current service time of the activity \emph{examine thoroughly} in the example process has an impact not only on the overall metrics but also on the duration of the later activities in the traces, e.g., \emph{decide} or \emph{reinitiate request}. 
\paragraph{Aggregated Performance Spectrum}
Performance Spectrum is a concept introduced to visualize the performance of process steps at the detailed level. A process segment in event log $L$ is a step from activity $a$ to activity $b$, i.e., $(a,b)\in A_L\times A_L$ is a process segment in $L$ where $A_L=\{ a \in \sigma | \sigma \in \widetilde{L}_{\mathcal{A}}\}$. 
%hand-over of work from resource a to b, or the movement of goods from location a to b. 
Each occurrence of a segment in a trace allows measuring the time between occurrences of $a$ and $b$ \cite{PerformanceSpecVadim2018}. %We first formalize the performance spectrum for a single process segment and then lift this to views on a process.
We define the set of all tuples of events that are directly followed in the traces in $L$ as $SEG^L=\{(e_i,e_{i+1}) | \exists_{\sigma = \langle e_1,e_2,...,e_n \rangle \in L}  e_i,e_{i+1} \in \sigma\}$. The projection of the events in $SEG^L$ on their activity attribute provides the multiset of process segments, i.e., $SEG^L_{\mathcal{A}}=[(\pi_{\mathcal{A}}(e_1),\pi_{\mathcal{A}}(e_{2})) | (e_1,e_2) \in SEG^L]$. For instance, $[(examine\ thoroughly, decide)^{17},(examine\ thoroughly, reinitiate\ request)^{20}]$ is the part of the multiset of segments in our example. %e.g., in $17$ of traces activity $examine\ thoroughly$ is directly followed by $decide$.

We consider two aspects for representing a process segment in an event log: \emph{average time} of the segment and \emph{frequency} of the segment.
%\autoref{fig:perfSpecExample} shows the similarity/differences between two processes (event logs) in terms of performance. 
%Each process segment is represented by a relative average service time and the frequency of the process segment. 
For $seg = (a,b)\in SEG_{\mathcal{A}}^L$, function $PS(seg, L)=(AvgTime(seg,L),Freq(seg,L))$ represents the frequency of the process segment $seg$ and the corresponding average time difference for the segment. %$t_b - t_a$. %average time between  
For $seg = (a,b)\in SEG_{\mathcal{A}}^L$, we define $AvgTime(seg,L) = Avg (\{\pi_{\mathcal{T}}(e_2)-\pi_{\mathcal{T}}(e_1)|(e_1,e_2)\in SEG^L \wedge \pi_{\mathcal{A}}(e_1){=} a \wedge \pi_{\mathcal{A}}(e_2){=}b \})$ and
$Freq (seg,L){=}SEG^L_{\mathcal{A}}((a,b))$. 

\begin{figure}[bt]
    \centering
    \includegraphics[width=\textwidth,]{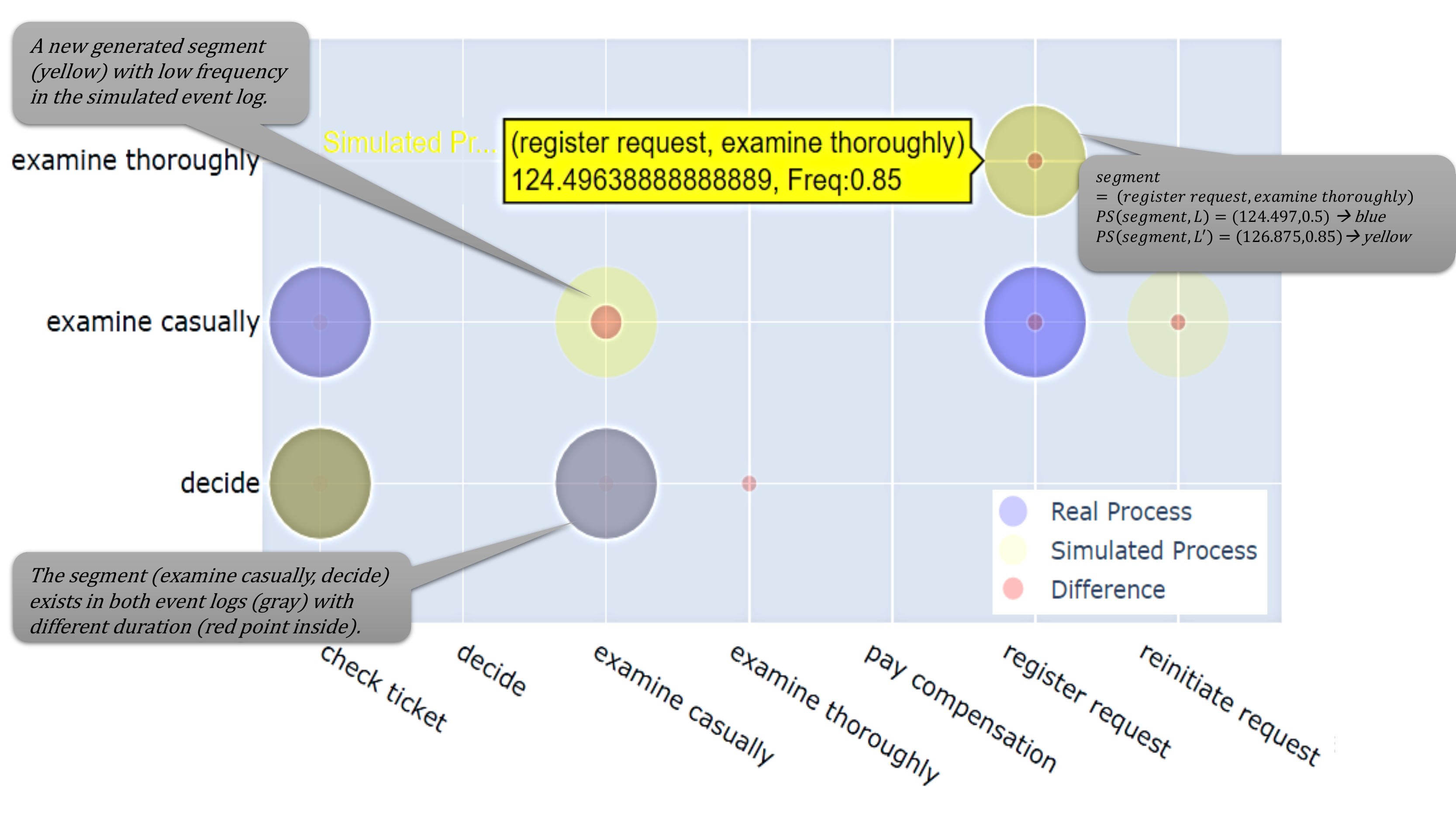}
    \caption{Part of the performance measurement for the example process based on the aggregated performance spectrum. Each event log is represented by a different color, i.e., blue for the original and yellow for the simulated one. 
Overlapping segments are represented by the gray color (same duration between segments).
Each point's transparency and size indicate the frequency and duration of the segment in the event logs. }
    \label{fig:perfSpecExample}
%\vspace{ -5 mm}
\end{figure}

\autoref{fig:perfSpecExample} is the result of the introduced performance measurement ($PS$) for the example process and the regenerated event log. In order to represent different aspects of the results, e.g., new/eliminated segments and different duration, we performed the simulation based on the changed process. For instance, given $L$ and $L'$ as the original and simulated event logs, each segment's colors refer to an event log, the size refers to the average time difference between the segments, and the transparency indicates the frequency (darker means more frequent). The gray color represents the overlapped segment in two event logs with similar performance metrics, and the yellow points represent the new segment generated in the simulated event log as a result of process tree choices. The implementation also includes the option to display only the difference (red points).
%and loops

\section{Evaluation}
\label{sec:eval}
A real event log representing the process of taking loans by customers inside a financial company, known as the \emph{BPI challenge 2012}, is used in this section. First, we simulate a similar process with different configurations and assess how close they are to the original event log. Following that, we alter the activity-flow of the process model in order to improve the process and evaluate the effect of the applied changes. %on the process. 
%design/improve a process model interactively in different situations and generate different event logs.  
Having both simulated and original behaviors of the process (with or without modifications) the possibility of comparing between two processes is easily provided. To do so, we used our tool \emph{SIMPT}\footnote{\scriptsize https://github.com/mbafrani/SIMPT-SimulatingProcessTrees} for simulating the process, and our developed modules for comparing two event logs w.r.t. the detailed performance and control flow aspects\footnote{\scriptsize https://github.com/mbafrani/VisualComparison2EventLogs}. The provided tools make it possible to evaluate the framework for the interactive improvement of different processes for different event logs.

\begin{figure}[bt]
    \centering
    \includegraphics[width=0.94\textwidth,]{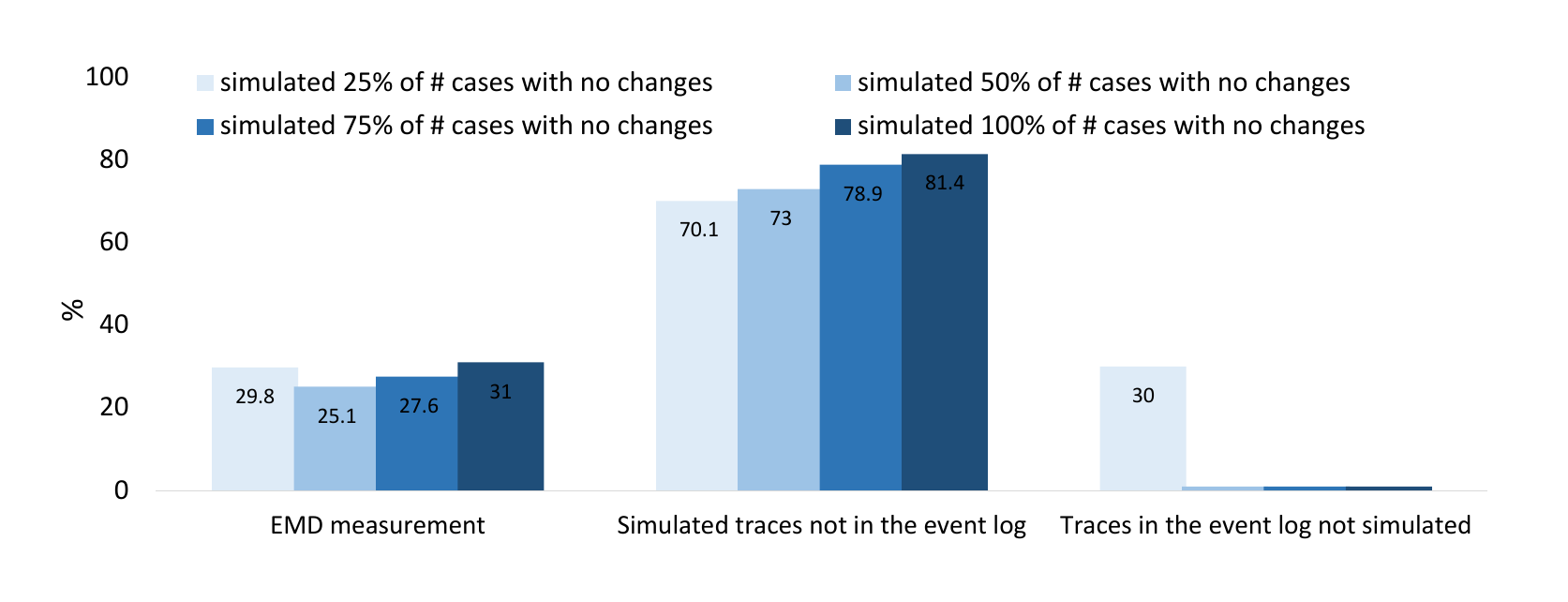}
    \caption{The comparison of the generated event logs using simulating a specific number of traces in the original event log (BPI Challenge 2012). The EMD measurement indicates, how the original and the simulated event logs are different. }
    \label{fig:emdEvalcahrt2012}
   % \vspace{-4mm}
\end{figure}

We start with automatically discovering and enriching the underlying process tree before regenerating the process, where the similarity of the two event logs indicates the possibility of using the simulation models for further investigation. Therefore, we simulated the event log multiple times without applying any changes. As shown in \autoref{fig:emdEvalcahrt2012}, we took a specific percentage of the total number of traces in the process for each round of simulation of the original process. As expected, when the number of simulated traces is small, there is a chance of missing specific process behaviors, e.g., using $25\%$ of the number of traces, we lost $30\%$ of the behaviors (unique traces). On the other hand, increasing the number of simulated traces increases the number of new behaviors. Since the generation of the traces (activity-flow) is based on probability and the process tree includes both $XOR$ choices and silent transitions, the new behaviors are expected to be generated.% simulation is based on the choices and silent transitions in the process tree  are the rerthe number of newly generated behavior increases.
\begin{figure}[bt]
    \centering
    \includegraphics[width=\textwidth, height= 0.21\textheight]{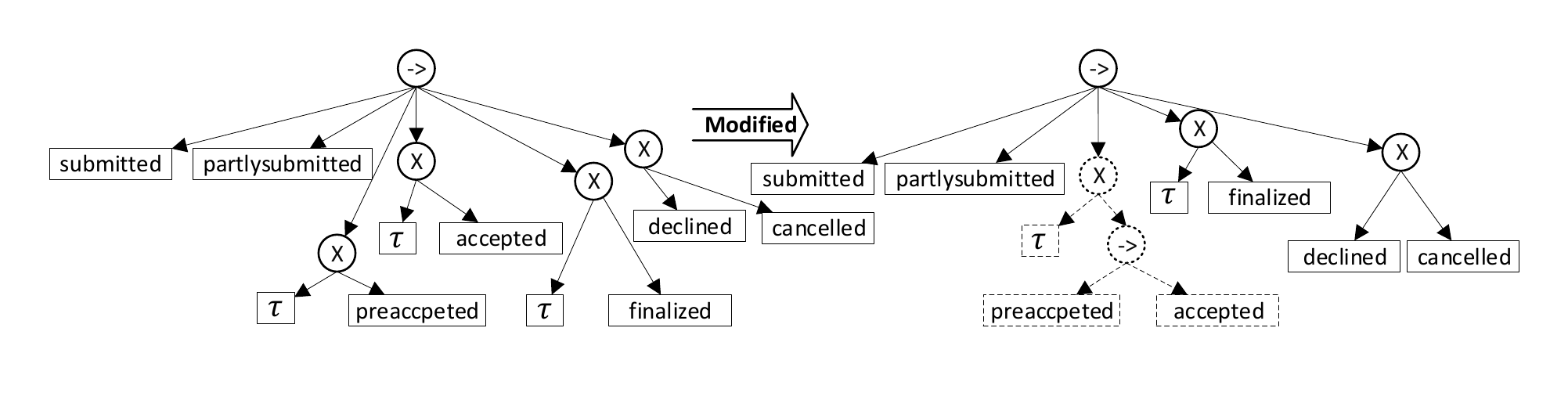}
    \caption{The process tree for handling application in the BPI challenge 2012 event logs (left). To evaluate the approach, the optional activity \emph{preaccepted} is changed to be mandatory in the flow of activities for all the traces (right) in the process. Dotted lines indicate the parts of the tree that have changed.  }
    \label{fig:eval2012ChangedTress}
\end{figure}
\begin{figure}[bt]
    \centering
    \includegraphics[width=\textwidth,height=0.22\textheight ]{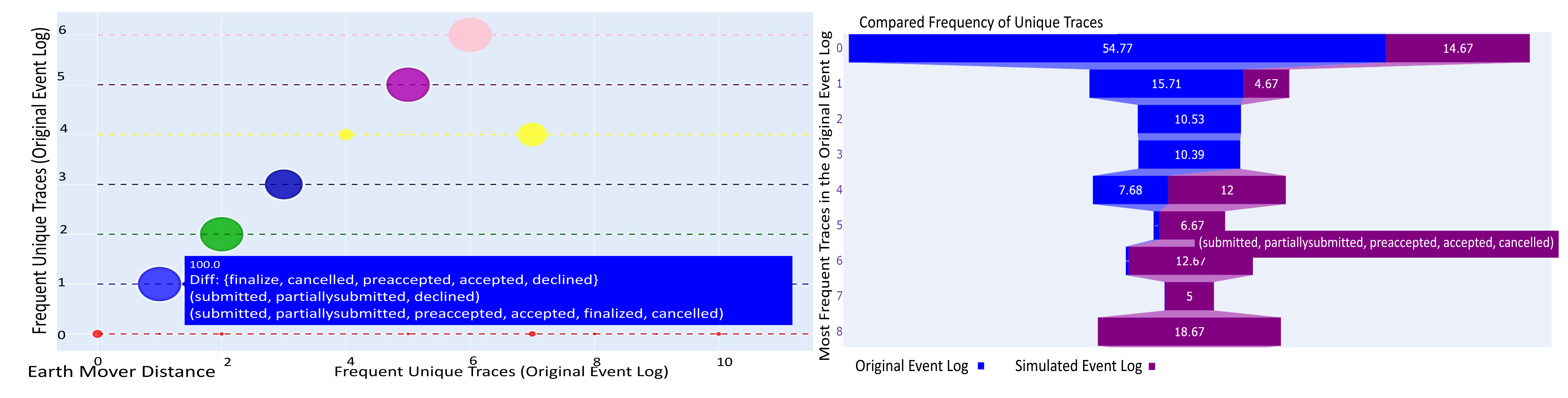}
    \caption{The detailed comparison of the changed process and the original process model. The detailed EMD diagram (left) shows the differences of the two event logs w.r.t. the activity-flow and the comparing frequent chart (right) represents the preserved and removed behavior in the simulated process as the effect of the applied changes.   }
    \label{fig:EvalComparemeasures2012}
   % \vspace{-5 mm}
\end{figure}

Afterward, in the process tree of the original process, we changed the optional activity \emph{preaccepted} to be a mandatory activity for all the traces that are going to be \emph{accepted} in \autoref{fig:eval2012ChangedTress}. The structure of the process tree (activity-flow) is changed from \small{ $\rightarrow( submitted, partlysubmitted, \times( \tau, preaccepted ), \times( \tau, accepted ),\\ \times( \tau, finalized ), \times( declined, cancelled ) )$ to $ \rightarrow( submitted , partlysubmitted , \times( \tau, \rightarrow( preaccepted , accepted )), \times( \tau, finalized ), \times( declined, cancelled ) )$}. \normalsize Note that these changes are possible in different aspects of the process such as the process model, performance metrics, e.g., activity duration, arrival rate of the traces, or capacity of the resources.  

Based on the shown results of simulating the original event log without any changes in \autoref{fig:EvalComparemeasures2012}, we simulated the changed process model with $50\%$ of original traces. In the proposed scenario (changed process tree), $63.6\%$ of generated behaviors (unique traces) are new. However, it is less than the behaviors in the simulated event log without any modifications, since we removed one of the $XOR$ choices limiting the possibilities of producing new behaviors. On the other hand, $23\%$ of the behaviors due to the change in the process tree are eliminated, i.e., the traces that skipped the activity \emph{preaccepted} in the original process. Also, in \autoref{fig:emdEvalcahrt2012}, the pairwise comparison of the traces (right), as well as the detailed EMD companions for the cost of the mapping of two event logs (left) after the changes, are shown. The applied changes in the process model not only affected the process behavior but also these changes affected the performance of the later segments in the process, e.g., the duration for the process segment \emph{accepted} and \emph{finalized} increased while activity \emph{finalized} was not changed. The provided detailed comparison along with the intermediate simulation tool enables the possibility of capturing these types of unexpected insights.
Note that the reliability of the simulation techniques such as the presented ones in \autoref{sec:rel} can be assessed using the measurement modules.

\section{Conclusion}
\label{sec:conclusion}
Process mining supports organizations in finding running processes, as well as identifying challenges or possible areas for improvement. The process improvement should be supported with process knowledge. We use process mining insights and simulation models as an intermediate method to regenerate processes in various scenarios. %Furthermore, assessing the impact of changes provides business owners with data-driven steps that can be taken to enhance their process. 
The framework begins with an event log, discovers a process tree, and enriches it with all the knowledge needed to regenerate the process. The similarity of the simulated results and the original process behavior in the form of an event log is then measured in the next step. The degree of similarity reflects the accuracy of our model. As a result, the improvement of the change in the process can be played out, and the impact of changes can be tracked using the same measurement module in both the activity-flow and performance aspects of the process. The advantage of our framework in both generating simulation models and enriching them based on event logs automatically, and the new representation of the comparing of the event logs. Furthermore, the intermediate simulation technique described in this paper can be replaced with other simulation techniques capable of generating event logs for the specified changes. 

\subsubsection*{Acknowledgments}  Funded by the Deutsche Forschungsgemeinschaft (DFG, German Research Foundation) under Germany's Excellence Strategy–EXC-2023 Internet of Production – 390621612. We also thank the Alexander von Humboldt (AvH) Stiftung for supporting our research.

\bibliographystyle{splncs04}
\bibliography{Reference}

\end{document}